\documentclass[aps,preprint,showpacs,nofootinbib,superscriptaddress]{revtex4}

\usepackage{latexsym}
\usepackage{amsmath,amsfonts}
\usepackage{amsbsy}
\usepackage{mathrsfs}
\usepackage{color}

\usepackage{psfrag}

\usepackage{enumerate}

\usepackage{amsmath,amssymb,calc,amsfonts}
\usepackage{latexsym}

\usepackage{comment}
\usepackage{tensind}
\tensordelimiter{?}

\newcounter{mnotecount}[section]

\usepackage{graphicx,calc,epsfig}
\def\ut#1{\rlap{\lower1ex\hbox{$\sim$}}#1{}}

\newcommand{\be}{\nopagebreak[3]\begin{equation}}
\newcommand{\ee}{\end{equation}}
\newcommand{\ba}{\nopagebreak[3]\begin{eqnarray}}
\newcommand{\ea}{\end{eqnarray}}

\DeclareFontFamily{U}{rsfs}{}         
\DeclareFontShape{U}{rsfs}{m}{n}{<5> rsfs5 <6><7> rsfs7          %
  <8><9><10><10.95><12><14.4><17.28><20.74><24.88> rsfs10}{}     %
\DeclareMathAlphabet{\mathfs}{U}{rsfs}{m}{n}                     %
\newcommand{\mfs}[1]{\mathfs {#1}}                               %
\newcommand{\inter}{{\lrcorner}}\newcommand{\n}{{\nonumber}}
\newcommand{\va}{\scriptscriptstyle}
\newcommand{\van}{\scriptstyle}

\newcommand{\sH}{{\mfs H}}
\newcommand{\sL}{{\mfs L}}

\newcommand{\sA}{{\mfs A}}

\def\pb#1{\rlap{\lower1.5ex\hbox{$\longleftarrow$}}{#1}}
\def\dpb#1{\rlap{\lower1.5ex\hbox{$\Longleftarrow$}}{#1}}
\def\spb#1{\rlap{\lower1.5ex\hbox{$\leftarrow$}}{#1}}
\def\sdpb#1{\rlap{\lower1.5ex\hbox{$\Leftarrow$}}{#1}}

\definecolor{blue}{rgb}{0,0,1}
\definecolor{green}{rgb}{0,1,0}
\definecolor{red}{rgb}{1,0,0}
\definecolor{vio}{rgb}{1,0,1}
\definecolor{ama}{rgb}{1,1,0}

\begin{document}



\title{Modelling black holes with angular momentum in loop quantum gravity}


\author{Ernesto Frodden}
\affiliation{Centre de Physique Th\'eorique\footnote{Unit\'e Mixte de Recherche (UMR 6207) du CNRS et Aix-Marseille Universit\'e; laboratoire affili\'e \`a la FRUMAM (FR 2291).}, Campus de Luminy, 13288 Marseille, France.}
\affiliation{P. Universidad Cat\'olica de Chile, Casilla 306, Santiago 22, Chile.}

\author{Alejandro Perez}
\affiliation{Centre de Physique Th\'eorique\footnote{Unit\'e Mixte de Recherche (UMR 6207) du CNRS et Aix-Marseille Universit\'e; laboratoire affili\'e \`a la FRUMAM (FR 2291).}, Campus de Luminy, 13288 Marseille, France.}

\author{Daniele Pranzetti}
\affiliation{Max Planck Institute for Gravitational Physics (AEI), Am M\"uhlenberg 1, D-14476 Golm, Germany}

\author{Christian R\"oken}
\affiliation{Centre de Physique Th\'eorique\footnote{Unit\'e Mixte de Recherche (UMR 6207) du CNRS et Aix-Marseille Universit\'e; laboratoire affili\'e \`a la FRUMAM (FR 2291).}, Campus de Luminy, 13288 Marseille, France.}

\vskip.5in


\begin{abstract} We construct a $SU(2)$ connection formulation of Kerr isolated horizons. As in the non-rotating case, the model is based on a $SU(2)$ Chern-Simons theory describing the degrees of freedom on the horizon.  The presence of a non-vanishing angular momentum modifies the admissibility conditions for spin network states. Physical states of the system are in correspondence with open intertwiners with total spin matching the angular momentum of the spacetime. 
\end{abstract}


\maketitle

\section{Introduction}

The phase space of rotating (Kerr) isolated horizons has  been characterized already in the very early papers on isolated horizons \cite{Lewandowski:2000nh}.
However, its quantization in the loop quantum gravity  framework has remained elusive due to what it seemed at first a technical issue:
as a result of the presence of angular momentum (a non-trivial charge generating rigid rotations around the symmetry axis) diffeomorphisms associated to vector fields
tangent to the horizons are not gauge symmetries of the system. 

Even though this breaking of some of the gauge symmetries by the boundary conditions has nothing 
pathological in itself and can be found in more familiar contexts
\footnote{Notice that this in strict analogy to the fact that generic diffeomorphisms that do not properly fall off at infinity are not gauge symmetries of the phase space of asymptotically flat solutions of general relativity.}, it introduces serious problems for the quantum theory if one tries to approach the issue of quantization using loop quantum gravity (LQG) techniques. The reason is that diffeomorphism invariance is at the heart of the definition of the LQG framework. Consequently, it can only 
accommodate boundary conditions that respect this fundamental symmetry. 

This is apparent from the central role played by diffeomorphism invariance 
in the models leading to the black hole entropy calculations for the Schwarzschild-type boundary condition. More precisely,  kinematical states of the spherically symmetric system are 
given by spin network states puncturing the horizon and endowing it with an area eigenvalue within the range $[A-\epsilon,A+\epsilon]$. The degeneracy of such kinematical states is infinite
as it is labelled by the coordinates defining the embedding of the punctures on the horizon. Physical states are however finitely many. The reason is that, according to the standard recipe of Dirac 
quantization, they are  
obtained by modding out gauge symmetries which in this case include tangent diffeomorphisms to the horizons. This is crucial for the finiteness of the entropy.  
This central step is not justified in the naive treatments of the rotating case. 
The lack of diffeomorphism invariance in the phase space of the Kerr isolated horizon 
makes the usual program inapplicable. 

An approach to deal with generic quantum isolated horizons (including rotation) has been proposed in \cite{Ashtekar:2004nd}. However, the question of the fate of the 
diffeomorphism symmetry is unclear in such treatment. In particular in such formulations both the leading order of the entropy calculation as well as the logarithmic corrections remain the same as the one of a non-rotating, spherically symmetric model. In this work we emphasize the central role of diffeomorphism invariance in the construction of the model of quantum rotating horizons. This will drastically change the nature of the admissible states to be counted in the entropy calculation, and will make them very natural from the perspective of previously stated intuitions about rotation in the context of LQG \cite{Krasnov:1998vc, Bojowald:2000pm}. As a consequence, the leading order term of the black hole entropy is not modified but the logarithmic corrections are, even in the special non-rotating case \cite{erka}. This resolves an apparent tension between different approaches to the problem of black hole entropy calculation \cite{Sen:2012dw}.

One can recover a manifestly diffeomorphism invariant description of the phase space of a rotating isolated horizon by 
appropriately including new degrees of freedom that restore the broken symmetry. This has been shown explicitly in \cite{Perez:2010pq}
using vector variables.
We will adapt the same idea to the connection variable formulation presented here. In fact what we aim at is a generalization of the Chern-Simons 
formulation used in the spherically symmetric context.

However, the first naive attempt to follow this strategy fails due to the fact that, in contrast to the spherically symmetric case, the pull-back to the horizon of the  Ashtekar-Barbero connection does not satisfy the simple boundary condition of the form $F(A)=(constant)\Sigma$, where  $\Sigma=e\wedge e$ \cite{Rocken:2012}. As this boundary condition becomes the key constraint equation for Chern-Simons theory in the non-rotating case this seems to rule out the possibility of describing the boundary degrees of freedom in terms of a Chern-Simons  theory in the rotating model.
 Additional heuristics that seems to preclude the Chern-Simons treatment of the rotating case comes from the
natural assumption, first put forward by Krasnov \cite{Krasnov:1998vc}, that quantum states of rotating horizons with total angular momentum $J$ should satisfy an additional constraint taking the form $J= \sum_{p} J_p$ (where $J_p$ are the spin operators associated to punctures of the horizon). In other words one assumes that 
the total angular momentum of the black hole is made up from microscopic contributions from individual spins in the punctures. This suggestion is certainly appealing from an intuitive perspective and from what we know about  LQG couplings to spinning matter, yet (with the exception of the symmetry reduced context \cite{Bojowald:2000pm}) it has not 
been established mathematically as far as we know.
Nevertheless, the point we want to make is that if such a constraint would be true then this would preclude the use of a Chern-Simons formulation as in such formulations one always obtains the closure constraint $\sum_{p} J_p=0$ from the equations of motion.

\begin{figure}
\psfrag{n}{$j_N=\frac{J}{2}$}
\psfrag{s}{$\!\!\!\!\!\!\!\!\!\!\!\!\!\!\!j_S=\frac{J}{2}$}
 \centerline{\hspace{0.5cm} \(
 \begin{array}{c}
\includegraphics[width=6.5cm]{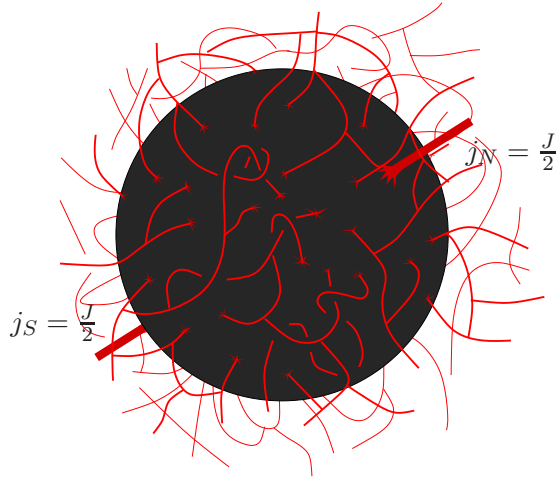}
\end{array}\) }
\caption{The admissible states of the rotating black hole are in correspondence with invariant vectors in the Chern-Simons Hilbert space 
$\sH^k_{\va CS}={\rm Inv}[j_1\otimes j_2\otimes \cdots j_n\otimes {J}]_k$ 
where $\{j_i\}_{i=1}^n$ are the spins carried by spin network punctures (with arbitrary $n$) and there are two additional (macroscopic) punctures at the south and north poles carrying spins $J/2$ respectively.
The subindex $k$ is there to remind one that the notion of invariant space are those of the quantum group $su(2)_q$ with $q$ fixed by the level $k$.}
\label{figure}
\end{figure}

The two apparent difficulties evoked in the previous paragraph are nicely avoided in one stroke as follows.
We will show that, using the available structure on the Kerr isolated horizon, one can introduce a new connection $\sA$ such that by definition one has 
$F(\sA)=(4\pi/k)\Sigma$ for $k$ constant almost everywhere on the horizon (we get to this key subtlety in a moment). 
If one uses $\sA$ as the connection dynamical field instead of $A$ then the boundary symplectic structure  takes the Chern-Simons form 
as far as the connection field is concerned. However, on the basis of our discussion in the previous paragraph,  this would seem to contradict Krasnov's natural intuition that the total spin contributed by the 
bulk geometry $\Sigma$ should be simply related to the spin of the black hole. In fact it does not. The reason is that the transformation from $A$ to $\sA$
produces singularities of $\sA$ in the north and south poles of the horizon as defined by the symmetry axis. The equation satisfied by the Cherns-Simons connection is
\be
\frac{k\ell_p^2}{4\pi}F(\sA)=\frac{\Sigma}{8\pi \gamma}+\frac{J}{2} \delta_N+\frac{J}{2} \delta_S, 
\ee  
where $J$ is the macroscopic angular momentum and the delta symbols represent singularities of the curvature at the north and south poles of the horizon
as defined by the singularities of the axisymmetric killing field. The previous constraint implies, in the quantum theory, that the total spin contribution of spin network punctures must 
add up to $J$ (modulo $k/2$). Admissible states can then be depicted as in Figure \ref{figure}. 

The geometric picture associated with the admissible states is similar to the one advocated in polymer models of the horizon geometry introduced in \cite{Bianchi:2010qd} and later in \cite{Livine:2012cv}.

\section{rotating horizons}

In this section we present the variables used in order to describe boundary degrees of freedom as Chern-Simons theory.
We will be able to show explicitly a classical solution in these variables such that isolated horizon conditions imply a consistent phase space description.
The pull-back of the Ashtekar-Barbero connection of Kerr geometry on the horizon has been computed in \cite{Rocken:2012}. 
Here we follow a different approach: instead of computing the pull-back of a bulk connection in Kerr geometry we construct a connection field $\sA^i$ 
from the Kerr horizon data. More precisely the Chern-Simons connection $\sA$ is required to satisfy the following set of conditions that will completely fix it up to gauge transformations and diffeomorphisms tangent to the horizon $H$. First we require the equation
\be
\frac{k}{4\pi} F^i(\sA)=\frac{1}{8\pi \gamma \ell^2_p}\Sigma^i,
\label{CCSS}
\ee
where $k$ is the Chern-Simons level which is a function of the area $A$ and the angular momentum $J$ of the isolated horizon that will be determined in what follows, to hold. 
The two-forms of the previous equation are pulled back to the horizon two-surface $H$.
The densitized triad field $\Sigma^i$ (the pull-back of $\epsilon^{ijk}e_i\wedge e_k$ to $H$, where $e_i$ is the co-tretrad field) is part of  the geometric data provided by the Kerr horizon geometry.

The above equation fixes the connection $\sA$ up to a rotation around the internal axis leaving $\Sigma^i$, seen as an internal vector, invariant. 
Explicitly, if $\sA_1$ is also a solution of (\ref{CCSS}) then $\sA_2=g\sA_1g^{-1}-gdg^{-1}$ is a solution of (\ref{CCSS}) with the same $\Sigma$ if  $g\in U(1)_{\Sigma}\subset SU(2)$ such that $g\Sigma g^{-1}=\Sigma$.
We view this as an intrinsic ambiguity in the choice of the variable $\sA$ and not a gauge transformation. In particular the bulk connection is (by definition) unaffected  by the transformation described above. Hence, we can and will exploit this freedom to fix our variable $\sA$ so that 
an additional condition is satisfied, namely 
\ba
v^{\perp}\inter(A^i-\sA^i)\Sigma_i=0,
\label{diff}
\ea 
where $v^{\perp}$ is the unique normal direction to the axisymmetric Killing field $\psi=\partial_{\varphi}$ on the horizon. In the usual spherical local coordinates the previous condition can simply be written as $(\partial_\theta)\inter(A^i-\sA^i)\Sigma_i=0$. We also require \be\label{lie}\sL_\psi(\sA^i\Sigma_i)=0,\ee
where, again, $\psi=\partial_\varphi$ is the axial Killing vector field on the Kerr horizon. With these conditions the connection $\sA$ is almost completely fixed by the data provided by $\Sigma$ and $A$ of the Kerr isolated horizon. The remaining freedom is fixed by the condition
\be\label{angu}
{J}=\frac{1}{8\pi\gamma}\int_{H}\psi\inter(A^i-\sA^i) \, \Sigma_i,
\ee
where $J$ is the total angular momentum of the spacetime.
Given $k$, equations (\ref{CCSS}) to (\ref{angu}) uniquely determine the connection $\sA^i$ up to gauge transformations and tangent diffeomorphisms (transforming $A,\Sigma$ and $\sA$). 

In order to study the properties of $\sA^i$  in more detail we will construct an explicit solution. The properties of this solution discussed below are all gauge and diffeomorphism 
invariant. We start with a type I connection $A_0^i$ (see Appendix in \cite{Engle:2010kt})
\ba && \n  {A_{0}^1}= \cos(\theta) d\phi \\
&& \n  {A_{0}^2}= \frac{1}{\sqrt{2}}( \sin(\theta) d\phi+\bar\gamma
d\theta) \\ 
&& \n  {A_{0}^3}= \frac{1}{\sqrt{2}}(\bar\gamma \sin(\theta)d\phi- d\theta). \ea 
The parameter $\bar\gamma$ is not determined for the moment. The previous connection will be used as a `seed' for constructing the Chern-Simons connection $\sA$ in what follows.
The fact that it is just the usual type I connection of \cite{Engle:2010kt} will guarantee that we recover the standard connection in the limit $J\to 0$.
The parameter $\bar\gamma$  labels a one-parameter family of suitable type I $SU(2)$ connections\footnote{This ambiguity exists in general. For a discussion see \cite{dpp}, and also the appendix in \cite{Rezende:2009sv},
where the ambiguity parameter controlling it is denoted by a dimensionful quantity $\lambda_2$. The type I geometry comes with a dimensionful scale (its area) and so the ambiguity  
becomes natural in such a context and can be labelled by a dimensionless parameter $\bar \gamma$.}.  In \cite{Engle:2010kt} the seemingly natural choice $\gamma=\bar\gamma$ was made.  We will see here that the inclusion of rotation gives us the means to fix this ambiguity in a more physical way by requiring that the level of the Chern-Simons theory (computed below) vanishes in the extremal case
$A=8\pi J$. The disappearance of the level in the extremal case will in turn imply that the entropy of an extremal black hole vanishes \cite{Hawking:1994ii}.
This will however have little effect on the entropy of physical black holes  no matter how close they are to the extremal case.

The curvature of the previous connection is
\be
F^i(A_0)=\delta^i_1 \frac{(\bar\gamma^2-1)}{2}\sin(\theta) d\theta\wedge d\varphi.
\ee
The solution that we are looking for can be obtained via an active diffeomorphism  $\phi_{W}$ acting on  $A_0$ 
sending $d\varphi\to \partial_\varphi W(\phi) d\varphi$.  Such action should not be confused with a gauge transformation as the diffeomorphism acts only on $A_0$. 
The action on the type I connection is $A_0\to \phi^*_{W} A_0$ and it follows immediately  that  
\be
F^i(\phi^*_{W} A_0)={\delta^i_1} \frac{(\bar\gamma^2-1)}{2}\sin(\theta)  \partial_\varphi W(\phi) d\theta\wedge d\varphi.
\ee
Now equation (\ref{CCSS}) becomes the following equation for $W(\varphi)$
\ba
&&{k}\, \partial_\varphi W(\varphi) = \frac{A}{4\pi \gamma {(\bar\gamma^2-1)}\ell^2_p}.
\ea
Thus $\phi^*_{W} A_0$ solves (\ref{CCSS}) if $W(\phi) =\frac{1}{k} \frac{A}{4\pi \gamma {(\bar\gamma^2-1)}\ell^2_p}\varphi$. The non-single-valued nature of $W(\varphi)$ will produce two curvature singularities at the poles. These will play a crucial role in the quantum theory. 

As mentioned above our connection has to fulfill also constraint (\ref{diff}) which is accomplished by fixing the $U(1)_{\va \Sigma}$ ambiguity. Considering all this our solution is given by
\be\label{solu}
\sA=g[\phi^*_{W} A_0]g^{-1}+gdg^{-1}
\ee
which is completely fixed (up to gauge transformations) by equations (\ref{CCSS}), (\ref{diff}), (\ref{lie}), and (\ref{angu})  and hence by the data contained in $A$ and $\Sigma$ for a Kerr isolated horizon.
Now, it is easy to show from (\ref{solu}) that in a circulation of an infinitesimal loop around the poles our variables satisfy
\be
\frac{k}{4\pi}\oint_{C}\sA^1
=\frac{A}{8\pi \gamma {(\bar\gamma^2-1)}\ell^2_p}.
\label{circu}
\ee 
The previous equation will be used to fix the value of the Chern-Simons level $k$. We require that
\be
\frac{k}{4\pi}\oint_{C}\sA^1
=\frac{k}{2}+\frac{J}{2\ell^2_p}.
\label{circu1}
\ee 
From equations (\ref{circu}) and (\ref{circu1}) we obtain $k={A}/({4\pi \gamma {(\bar\gamma^2-1)}\ell^2_p}) - {J}/{\ell_p^2}$.
 The level of the Chern-Simons connection is given by the usual non-rotating level minus the isolated horizon angular momentum in Planck units.
 We choose to fix the ambiguity parameter $\bar\gamma=\sqrt{(2+\gamma)/\gamma}$ so that the Chern-Simons level takes the simpler form\footnote{Notice that one could fix $\bar\gamma$ so that  \be k=\frac{\sqrt{A^2-(4 \pi q)^2}}{8\pi \ell_p^2}-\frac{J}{2\ell^2_p}.\ee This would imply that $k$ vanishes for all possible extremal horizons \cite{Dain:2011pi, Jaramillo:2011pg, Dain:2011kb, Clement:2012vb}.}
 \be
k=\frac{A}{8\pi\ell^2_p} - \frac{J}{\ell_p^2}
 \ee
 which has the important property that it vanishes in the extremal case $A=8\pi J$. We will comment further on the importance of this choice.

Equation (\ref{circu1}) implies the presence of conical singularities in the curvature $F^{i}(\sA)$ at the poles. 
We will see in the following section that these singularities are relevant for the implementation of the Chern-Simons quantization of the rotating isolated horizon.
One can recall the presence of the singularities at the poles if one writes the curvature equation over $H$ in its entirety (including the poles) as
\be
\frac{k}{4\pi} F(\sA)^i= \frac{\Sigma^i}{8\pi  \ell_p^2 \gamma}+p \delta^i_1 \delta_{\va N} +p \delta^i_1\delta_{\va S}, 
\label{singu}
\ee
where $\delta_{\va N}$ and $\delta_{\va S}$ are Dirac delta functions centred on the north and south poles, respectively, and
\ba
p=\frac{k}{2}+ \frac{J}{2\ell_p^2} \label{pes}.
\ea
In the quantum theory we will see that $p$ appears in a quantum constraint which, due to the 
properties of quantum Chern-Simons theories, will be sensitive to $p$ modulo $\frac{k}{2}$, here denoted by $[p]_{\frac{k}{2}}$. Therefore, we have
\be
[p]_{\frac{k}{2}}=\left[\frac{J}{2\ell_p^2}\right]_{\frac{k}{2}}.
\ee 

\noindent{\em Remark:} There is a non-trivial choice in equation (\ref{circu1}) that determines the value of the Chern-Simons level. This choice implies that quantum states of the rotating
horizon are given by vectors in the representation of rigid gauge transformations with total angular momentum $J=\int_{\va H} j$, where $j$ is the angular momentum density of 
the rotating horizon that will be explicitly introduced in the following section. This interpretation is available at least when $J\le {k}$. For other values of $J$, allowed by the 
classical inequality $8\pi J\le A$, the interpretation is less obvious as the quantum group structure coming from the quantization of Chern-Simons theory becomes relevant. 
We will see in a more extended discussion in Section \ref{conclu} that this feature eliminates some apparently puzzling inconsistencies with the classical black hole properties found in \cite{Krasnov:1998vc}. 
If we would have replaced the right hand side of equation (\ref{circu1}) by $\frac{k}{2}$, the conical singularities of the
connection at the poles would not have had an effect at the quantum level and physical states would be invariant vectors (intertwiners) under rigid gauge transformations. The latter choice corresponds to the ($SU(2)$) generalization of the type I connection technology used 
in \cite{Ashtekar:2004nd}. This second option is logically possible and one cannot rule it out on the basis of first principles. The strength of the choice made here is that it produces quantum states with a simple geometric interpretation. It leads to a Chern-Simons level that vanishes in the extremal case, and modifies the logarithmic corrections to the entropy computation.

\section{Conservation of the symplectic structure}

In this section we present the symplectic structure and prove that it is conserved provided that the standard boundary conditions hold. The 
symplectic structure is constructed in terms of the connection $\sA$ introduced in the previous section. Additional variables
are necessary to preserve diffeomorphism invariance in the rotating case (see \cite{Perez:2010pq} for a discussion). These are a two form $j$ (that will acquire the physical meaning of the angular momentum density on shell) 
and its conjugate momentum, a scalar field $\Phi$.  

As in the usual treatment \cite{Ashtekar:1997yu} the only allowed variations on the horizon are tangent diffeomorphisms and $SU(2)$ gauge transformations.
We start with the $SU(2)$ gauge transformations denoted by $\delta_{\alpha}$ for $\alpha(x)\in su(2)$, i.e a Lie algebra valued scalar on $M$. For the bulk variables we have
\ba \label{gaugM}
&& \delta_{\alpha} \Sigma=[\alpha,\Sigma] \n \\
&& \delta_{\alpha} A=-d_A\alpha,
\ea 
while for boundary variables the transformation is
\ba \label{gaugB}
&& \delta_{\alpha} \sA =-d_\sA\alpha \n \\
&& \delta_{\alpha} \Phi=({\alpha_1|_{\va N}+\alpha_1|_{\va S}})/{2} \n \\
&& \delta_{\alpha} j=0.
\ea 
Note that the angular momentum density $j$ is gauge invariant by construction and the 
scalar field transforms in a distributional way: only the value of $\alpha$ on $H$ at the symmetry axis (the north and south poles)
change $\Phi$. 

We restrict diffeomorphisms to vector fields $v$ that vanish at the north and south poles of $H$ and, therefore, leave the north and south poles invariant.
The transformation $\delta_v$ is 
\ba\label{diffy}
&& \delta_v \Sigma=\sL_v\Sigma= d(v\inter \Sigma) \n \\
&&\delta_v A=\sL_v A= v\inter dA+ d(v\inter A)\n \\
&&\delta_v \sA=\sL_v\sA=v\inter d\sA+ d(v\inter \sA)\n \\
&&\delta_v j=\sL_vj= d(v\inter j)\n \\
&&\delta_v\Phi=v\inter d\Phi.
\ea

\noindent{\bf Proposition:}  In terms of the Ashtekar-Barbero variables the presymplectic structure
of the rotating Kerr horizon takes the form
\ba
\label{BIsylstr} \n \!\!\!\!\!\!\!\!\!\!\!\!\!\Omega_{M}&=&\Omega_{\va B}+\Omega_{\va H}\\ &=&\frac{1}{\kappa\gamma }\int_{M}\!\!\! 2\delta_{[1}
\Sigma^{i} \wedge \delta_{2]} A^{\va}_{i} + {\frac{k}{4 \pi}} \int_H \!\!\! \delta_{1} \sA_i \wedge \delta_2
 \sA^i-\frac{16\pi}{\kappa} \int_{H} \delta_{[1}\Phi \, \delta_{2]} j, 
\ea
where $k$ is the level of the CS boundary term and $\kappa=8\pi G$. $\Omega_{\va B}$ denotes the first (bulk integral) term while $\Omega_H$ denotes the last two (surface integral) terms. 

\vskip.2cm
\noindent{\em Proof:} We prove the result by first looking at variations which are pure $SU(2)$ gauge transformations. Then we show the invariance for pure diffeomorphisms.

\begin{center}
{\em Invariance under infinitesimal $SU(2)$ transformations}
\end{center}

We want to check that $\Omega_{M}(\delta_{\alpha}, \delta)=\Omega_{\va B}(\delta_{\alpha}, \delta)+\Omega_{\va H}(\delta_{\alpha}, \delta)=0$ for $\delta_{\alpha}$ which is a local $SU(2)$ transformation as given in (\ref{gaugM}) and (\ref{gaugB}).
The first contribution $\Omega_{\va B}(\delta_{\alpha},\delta)$ yields
\begin{eqnarray*}&&
  \Omega_{\va B}(\delta_{\alpha},\delta)\!=\!
 \frac{1}{\kappa \gamma}\int_{\va M}\!\!\! \left( [\alpha,\Sigma]_i \wedge \delta A^i \!+ \!\delta \Sigma_i \wedge d_A \alpha^i \right)
\\ &&
\!=- \frac{1}{\kappa \gamma} \int_{\va M}\!\!\! \left[ d (\alpha_i \delta\Sigma^i)\! - \!\alpha_i
\delta(d_A \Sigma^i)\!\right] =\! 
- \frac{1}{\kappa \gamma} \int_{\va H} \alpha_i \delta \Sigma^i,
\end{eqnarray*} where we have used the Gauss law
$\delta (d_A\Sigma)=0$ and that boundary terms at infinity vanish. At the boundary itself we have to take special care of the singular nature of our connection variables at the poles. Therefore, we split $H$ in two infinitesimal patches around the poles N and S, and an intermediate strip $H^*=H\backslash(N\cup S)$. Thus we obtain
\begin{eqnarray*}
\frac{k}{4\pi} \int_{H} \delta_{\alpha} \sA_i \wedge \delta\sA^i 
\!&=&\!  -\frac{k}{4\pi} \int_{H} d_{\sA}\alpha^i \wedge \delta\sA_i \\
\!&=&\!  -\frac{k}{4\pi} \int_{H^*} d(\alpha^i\delta \sA_i)+\frac{k}{4\pi} \int_{H^*} \alpha_i\delta F^i(\sA)-\frac{k}{4\pi} \int_{N\cup S} (d\alpha^i+?\epsilon^i_jk?\sA^j\alpha^k) \wedge \delta\sA_i\\
\!&=&\!  -\frac{k}{4\pi} \int_{\partial H^*}\alpha^i\delta \sA_i+\frac{1}{\kappa\gamma} \int_{H^*} \alpha^i\delta  \Sigma_i\\
\!&=&\!  \frac{k}{4\pi} \int_{\partial N}\alpha^i\delta \sA_i+\frac{k}{4\pi} \int_{\partial S}\alpha^i\delta \sA_i+\frac{1}{\kappa\gamma} \int_{H} \alpha^i\delta  \Sigma_i\\  
\!&=&\!  \int_{H} \alpha_i\delta\left( \frac{1}{\kappa\gamma}\Sigma^i +p \delta_{\va N}\delta_{1}^i+ p \delta_{\va S}\delta_{1}^i\right),
\end{eqnarray*}
where on line 2 we have integrated by part, on line 3 we used (\ref{singu}) on $H^*$, on line 4 we used $\partial H^*=-(\partial N\cup \partial S)$, and on line 5 we used (\ref{circu1}).
Then
\ba
 \Omega_{H}(\delta_{\alpha},\delta)\!&=&\!  \frac{k}{4\pi} \int_{H} \delta_{\alpha} \sA_i \wedge \delta\sA^i-\frac{8\pi}{\kappa} \int_{H} \delta_\alpha \Phi \, \delta j\\
\!&=&\!\frac{1}{\kappa\gamma} \int_{H} \alpha^i\delta  \Sigma_i+  (\alpha_1|_{\va N}+\alpha_1|_{\va S})\delta p-\frac{4\pi}{\kappa}(\alpha_1|_{\va N}+\alpha_1|_{\va N})\int_{H} \delta j
\label{gau}
\ea
Hence, the symplectic structure is gauge invariant, namely $\Omega_{M}(\delta_{\alpha}, \delta)=\Omega_{\va B}(\delta_{\alpha}, \delta)+\Omega_{\va H}(\delta_{\alpha}, \delta)=0$, if the following constraint is satisfied
\ba
\frac{8\pi}{\kappa} \int_{H} j&=& p. 
 \label{S}
\ea

\begin{center}
{\em Invariance under infinitesimal diffeomorphisms}
\end{center}

Now we focus on the invariance under infinitesimal diffeomorphisms, in other words, we want to show that for a small tangent vector field $v\in T(H)$ we have
$$\Omega_{M}(\delta_{v}, \delta)=\Omega_{\va B}(\delta_{v}, \delta)+\Omega_{\va H}(\delta_{v}, \delta)=0.$$ For the bulk term, using (\ref{diffy}), we obtain
\ba 
 \nonumber  \Omega_{\va B}(\delta_v,\delta) &=& \frac{1}{\kappa \gamma}\int_{M}
 \left[\sL_v\Sigma_i\wedge \delta A^{i}-\delta \Sigma_i\wedge \sL_vA^i\right]\\ \nonumber 
&=&\frac{1}{\kappa \gamma}\int_{M} \left[d_A(v\inter\Sigma)_i\wedge \delta A^{i}-\delta\Sigma_i\wedge v\inter F^i+d(v\inter A_i \, \delta\Sigma^i)\right]\\ \nonumber 
&=&\frac{1}{\kappa \gamma} \int_{M} \left[d(v\inter\Sigma_i\wedge \delta A^{i}) +v\inter\Sigma_i\wedge d_A(\delta A^{i})-\delta
 \Sigma_i\wedge v\inter F^i+d(v\inter A_i \, \delta\Sigma^i)\right]\\  \nonumber 
&=&\frac{1}{\kappa \gamma}\int_{M}\left[ d(v\inter\Sigma_i\wedge \delta A^{i})+v\inter\Sigma_i\wedge \delta F^{i}-\delta \Sigma_i\wedge v\inter F^i+d(v\inter A_i \, \delta\Sigma^i)\right]\\ \n 
&=&\frac{1}{\kappa \gamma}\int_{M}\left[ d(v\inter\Sigma_i\wedge \delta A^{i}) + \delta (\Sigma_i \wedge v\inter F^i(A))+d(v\inter A_i \, \delta\Sigma^i)\right] \\ &=&\frac{1}{\kappa \gamma} \int_{H} \delta (v\inter A_i \, \Sigma^i). \ea

The horizon term yields
\ba \nonumber
\Omega_{\va H}(\delta_{v}, \delta)&=& \frac{{k}}{4\pi } \int_{H} \sL_v \sA^i\wedge \delta \sA_{i} -\frac{8 \pi}{\kappa}\int_{H}\left[ \sL_v\Phi \, \delta j-\delta\Phi \, \sL_v j\right]\\ \nonumber 
&=&-\frac{k}{4\pi }  \int_{H}  [\delta \sA_{i}\wedge v\inter F^i(\sA)+\delta \sA_{i}\wedge d_{\sA} (v\inter \sA^i)]
-\frac{8 \pi}{\kappa} \int_{H} \left[v\inter d\Phi\, \delta j-\delta\Phi\, d(v\inter j)\right]\\ 
&=&-\frac{k}{4\pi }  \int_{H}  [\delta (v\inter \sA_{i})\, F^i(\sA)+\delta F_i(\sA)\, v\inter \sA^i]- \frac{8 \pi}{\kappa} \int_{H}\left[ v\inter d\Phi\, \delta j +\delta (d\Phi)\wedge v\inter j\right]\nonumber \\ 
&=& - \frac{k}{4\pi}  \int_{H}  \delta (v\inter \sA_{i}\, F^i(\sA)) - \frac{8 \pi}{\kappa} \int_{H} \delta (v\inter d\Phi\, j) \nonumber \\ 
&=&-\frac{1}{\kappa \gamma}\int_{H}\delta[ v\inter\sA^i\, \Sigma_i+ {8 \pi\gamma }\ v\inter d\Phi\, j]. 
\ea
Now, equation $\Omega_{\va M}(\delta_{v}, \delta) = 0$ is satisfied if the following constraint holds
\ba\label{vc}
\frac{1}{\kappa \gamma} \int_{H}\delta[ v\inter(A^i-\sA^i)\, \Sigma_i-8 \pi  \gamma \ v\inter d\Phi\, j] = 0
\ea
for all $v\in T(H)$. Equation (\ref{vc}) is nothing else but the diffeomorphism constraint in these variables.
The classical solution corresponding to Kerr is $\Phi=\varphi$, where $\varphi$ is the Killing parameter associated to axisymmetry.
In this case $(A^i-\sA^i)_{\varphi} \Sigma_i/( 8\pi \gamma)$ is the angular momentum density satisfying  
\be
{J}= \int_{H} j=\frac{1}{8\pi\gamma}\int_{H}(A^i-\sA^i)_{\varphi} \, \Sigma_i,
\ee
where $J$ is the total angular momentum of the Kerr solution.
This provides the physical interpretation of the $l.h.s$ of the constraint (\ref{S}) found above telling us that $p=8\pi J/\kappa$. 

\section{Quantization}

 Once the degrees of freedom on the boundary are captured by a Chern-Simons 
symplectic structure plus Chern-Simons-like constraint, as the one given in equation (\ref{singu}), the quantization is basically analogous to the one applied in the non-rotating case. 
There are, however, new aspects here that have to be treated carefully. The most obvious one is that in addition to the Chern-Simons connection
$\sA$ we have the field $j$ and its conjugate $\Phi$ in the boundary symplectic structure and their quantization too needs to be 
addressed. The second issue is that the Chern-Simons constraint (\ref{singu}) contains two classical singularities at the north and south poles of the sphere and these 
are seemingly new features of the rotating system.
Here we will start by ignoring the first problem and go directly to the second. The last part of this section will be dedicated to the first.
 
 { As in the non-rotating case, and if for the moment we concentrate on the connection fields, the form of the symplectic structure motivates one to handle the
quantization of the bulk and horizon degrees of freedom 
separately.
We first discuss the bulk quantization. As in standard LQG [8] one
first considers (bulk) Hilbert spaces $\sH^B_\gamma$ defined on a
graph $\gamma \subset M$ and then takes the projective limit
containing the Hilbert spaces for arbitrary graphs. Along these
lines let us first consider $\sH^{\va B}_{\gamma}$ for a fixed graph
$\gamma \subset M$ with end points on $H$, denoted $\gamma\cap H$.
The quantum operator associated with $\Sigma$ in (\ref{singu}) reads
%
%
\begin{equation}
\label{gammasigma} \epsilon^{ab}\hat{\Sigma}^i_{ab}(x) = 16 \pi G
\gamma \sum_{p \in \gamma\cap H} \delta(x,x_p) \hat{J}^i(p),
\end{equation}
where $[\hat{J}^i(p),\hat{J}^j(p)]=\epsilon^{ij}_{\ \ k} \hat{J}^k(p)$ at each $p\in\gamma\cap H$.
%
%
Also, consider a basis of $\sH^{{\va B}}_{\gamma}$ of eigenstates of both
$J_p\cdot J_p$ as well as $J_p^3$ for all $p\in \gamma\cap H$ with
eigenvalues $\hbar^2 j_p(j_p+1)$ and $\hbar m_p$, respectively. These
states are spin network states, here denoted by $|\{j_p,m_p\}_{\va
1}^{\va n}; {\van \cdots} \rangle$, where $j_p$ and $m_p$ are the
spins and magnetic numbers labelling the $n$ edges puncturing the horizon
at points $x_p$ (other labels are left implicit). They are
eigenstates of the horizon area operator $\hat a_{\va H}$ as well
\[ \hat a_{\va H}|\{j_p,m_p\}_{\va 1}^{\va
n}; {\van \cdots} \rangle=8\pi\gamma \ell_p^2 \,
\sum_{p=1}^{n}\sqrt{j_p(j_p+1)} |\{j_p,m_p\}_{\va 1}^{\va
n}; {\van \cdots} \rangle. \]

Now substituting the expression (\ref{gammasigma})
 into the quantum version of (\ref{singu}), we obtain
\begin{equation}\label{seven}
\frac{k}{8\pi}\epsilon^{ab}\hat{F}^i_{ab} =   \sum_{p \in \gamma\cap H} \delta(x,x_p) \hat{J}^i(p)-\delta(x,x_{\va N}) \ J_{ N}^i- \delta(x,x_{\va S}) \ J_{ S}^i,
\end{equation}
where 
\be\label{clas}
J_{ N}^i=\frac{J}{2\ell_p^2} \hat z^i \ \ \ \ {\rm and}\ \ \ \ J_{S}^i=\frac{J}{2\ell_p^2} \hat z^i
\ee
for $\hat z^i$ a normalized internal direction representing the symmetry axis.
As we will show, the previous equation tells us that the surface Hilbert space $\sH^{{\va
H}}_{\gamma\cap H}$ that we are looking for is precisely the one
corresponding to (the well studied) Chern-Simons theory in the presence of particles. Equation (\ref{seven}) implies the {\em formal} closure constraints
\ba\label{31}
 && \sum_{p \in \gamma\cap H} \hat{J}^z(p)=\frac{J}{\ell_p^2},\n \\
 && \sum_{p \in \gamma\cap H} \hat{J}^y(p)=0, \n \\
&&  \sum_{p \in \gamma\cap H} \hat{J}^x(p)=0.
\ea
We call them {\em formal} because they are indeed inconsistent due to quantum uncertainties. 
However, there is a clear consistent quantum version of the previous conditions. 

From the point of view of quantum geometry
(bulk perspective), admissible states (solving the above constraint in the strongest possible way compatible with the uncertainty principle)  are coherent states of the collection of punctures satisfying the constraints:
\ba\label{31}
&&\sum_p m(p)={\left[\bar J \right]_{k}} \label{jj} \n \\
&& \left[\sum_p J^{i}(p)\right]\left[\sum_p J_{i}(p)\right]=[{\bar J(\bar J+1)}]_{q(k)},
\ea
where $-j(p)\le m(p)\le j(p)$ denote the usual magnetic quantum numbers, $\bar J={J}/{\ell_p^2}$, and in the last equality, the r.h.s. denotes the $SU(2)_{q(k)}$ Casimir.
The state is of  the form $|\bar J,\bar J\rangle$ in the usual Wigner notation $|j,m\rangle$. Such states can be graphically represented as shown in Figure \ref{figure}. 

From the point of view of the boundary Chern-Simons theory the constraints are even simpler.
The two classical punctures are aligned along the same axis. This amounts in the Chern-Simons description 
to a single puncture carrying the total macroscopic spin of the black hole. This is the view taken in \cite{erka}. 
Admissible states span the intertwiner space $j_{1}\otimes j_2\otimes\cdots\otimes j_n\to \bar J$, give condition
(\ref{jj}), and finally the usual area constraint
\be
A-\epsilon\le8\pi\gamma\ell_p^2\sum_{p=1}^{n}\sqrt{j_p(j_p+1)}\le A+\epsilon.
\ee
 It can be seen that the leading order contribution of the entropy is not affected, yet logarithmic corrections are. A detailed calculation will be presented in \cite{erka}.

Finally we need to address the quantization of $j$ and $\Phi$ and the imposition of the vector constraint (\ref{vc}), namely
\ba\n
\int_{H}\delta[ v\inter(A^i-\sA^i)\, \Sigma_i-8 \pi   \ v\inter d\Phi\, j] = 0
\ea
for all vector fields $v$ tangent to $H$. 
At the classical level the previous constraint completely reduces the $(j,\Phi)$ degrees of freedom. This is due to the fact that it is an additional first class local constraint for two local degrees of freedom.
More precisely this constraint is responsible for imposing  diffeomorphism invariance. Here we assume that this holds also at the quantum level: for each spin network state 
satisfying the above restrictions there is only one solution of the previous equation for the quantum counterpart of $j$ and $\Phi$. In other words admissible states are indeed labelled by the spin quantum numbers satisfying the above constraints up to diffeomorphisms.  This assumption is similar to the one made generically in the context of quantum states of isolated horizons as far as the bulk Hamiltonian and diffeomorphism constraints are concerned. 
It would certainly be worth to be eliminated and it is probably within the reach of present background independent quantization techniques.

\section{Conclusions}\label{conclu}

In this work we have constructed a model of a rotating isolated horizon 
which is axisymmetric and has angular momentum $J$. The classical description of the system 
is based on a $SU(2)$ Chern-Simons connection plus additional auxiliary fields that restore 
diffeomorphism invariance. In the quantum theory the 
connection is constrained to be flat almost everywhere. As in spherically symmetric models, there are conical singularities with a 
strength that matches the quantum flux of the area encoded in the spin quantum numbers of spin network edges ending at the horizon.
In addition to these, there are two conical singularities at the north and south poles (as defined by the singularities of the axisymmetric Killing field) 
with combined strength equal to $[J/\hbar]_{k/2}$.
 
An ambiguity parameter in the definition of the $SU(2)$ boundary Chern-Simons connection, 
identified in previous models, can be fixed in the rotating case by the requirement that the level of the Chern-Simons theory 
vanishes in the extremal case. This requirement implies that the number of states of an extremal horizon is unity and hence that their
 entropy vanishes as suggested in \cite{Hawking:1994ii}. This is by no means in contradiction with the Hawking area law for non-extremal black holes: it can be shown \cite{erka} that the number of states of
 the rotating isolated horizon grows exponentially with the area with a universal coefficient (the same as in the non-rotating case) for large black holes no matter how close to the extreme case they are.
 Therefore, the entropy of physical black holes is consistent with the Hawking-Bekenstein entropy formula. The 
 proportionality constant is not, as in previous models, equal to $1/4$ (for a newly introduced perspective on the origin of the mismatch see \cite{Ghosh:2011fc}, and for an argument as to how the
 low-energy Bekenstein-Hawking entropy is to be recovered see \cite{Ghosh:2012wq}).
 
 Moreover, as shown in \cite{erka}, the logarithmic correction of the entropy for isolated horizons is corrected by the inclusion of angular momentum even in the non rotating-case. Corrections are universal 
 of the form $-2 \log(A)$. This is consistent with the results obtained using other methods \cite{Sen:2012dw} as far as non-local corrections are concerned. Local logarithmic corrections can arise from radiative corrections.
 In the context of the LQG framework a natural scenario for these corrections to appear is presented in  \cite{Ghosh:2012wq}. 
 
 In \cite{Krasnov:1998vc} a tension was pointed out between the analog of 
 equations (\ref{31}), the area spectrum of LQG, and the fact that classically $J$ can vary between $0$ and $A/(8\pi)$, which completely disappears in our formalism.
 In that reference the analogue of (\ref{31}) was postulated with the important difference that the r.h.s. would not contain the ${\rm mod}_k$ symbol.
 In such a case one sees that there are maximum spin states of the horizon for which $J_{max}\approx A/(8\pi\gamma)$. The fact that, classically, $J_{max}=A/(8\pi)$
 would seem to imply $\gamma=1$. Moreover, as the spectrum of the area is non-linear in the spins, it was conjectured in 
 \cite{Krasnov:1998vc} that the extremal black holes would be represented by single puncture states with a large spin: in the large spin limit the spectrum becomes linear.
 None of these conclusions are valid in our model due to the appearance of the symbol ${\rm mod}_k$ on the r.h.s. 
 Indeed any classically allowed angular momentum value leads to a consistent set of constraints and there are no restrictions on the
 value of the Immirzi parameter $\gamma$. No matter how close we are from the extremal situation the black hole states that dominate the statistical mechanical treatment have many punctures (of the order of $A/\ell_p^2$) which is compatible with the idea that these states approximate continuum geometries well.

\section{Acknowledgement}
CR was supported by a DFG Research Fellowship. EF was supported by CONICYT (Chile) grant D-21080187 and by the interchange program of BECAS-CHILE.

\end{document}